\documentclass[10pt]{article}
\usepackage{jheppub}
\usepackage{enumerate}
\usepackage{bm}
\usepackage{bbm}
\usepackage[usenames,dvipsnames]{xcolor}

\usepackage{amssymb,amsmath}
\usepackage{graphicx}
\usepackage{overpic}
\usepackage{color}
\usepackage[colorlinks=true]{hyperref}
\usepackage{enumitem}
\usepackage{braket}

\def\ba{\begin{eqnarray}}
\def\ea{\end{eqnarray}}
\newcommand{\R}{{R}}
\newcommand{\tT}{\text{T}}
\newcommand{\tR}{{\cal R}}

\newcommand{\scH}{\mathcal{H}}
\newcommand{\K}{\mathcal{K}}
\newcommand{\D}{\text {D}}
\newcommand{\G}{{\cal G}}
\newcommand{\g}{\bar{g}}

\def\d{\mathrm{d}}
\def\mn{_{\mu \nu}}
\def\({\left(}
\def\){\right)}
\def\p{\partial}

\begin{document}

\title{Perturbations of Stealth Black Holes in DHOST Theories}

\author[a,b]{Claudia de Rham}
\author[a]{\& Jun Zhang}
\affiliation[a]{Theoretical Physics, Blackett Laboratory, Imperial College, London, SW7 2AZ, UK}
\affiliation[b]{CERCA, Department of Physics, Case Western Reserve University, 10900 Euclid Ave, Cleveland, OH 44106, USA}

\emailAdd{c.de-rham@imperial.ac.uk}
\emailAdd{jun.zhang@imperial.ac.uk}

\abstract{Among the Scalar--Tensor modified theories of gravity, DHOST models could play a special role for dark energy while being consistent with current observations, notably those constraining the speed of gravitational waves. Schwarzschild-de Sitter black holes were shown to be exact solutions of a particular subclass of quadratic DHOST theories, while carrying a nontrivial scalar profile that linearly evolves in time and hence potentially providing exciting new phenomenological windows to explore this model. We investigate the physical perturbations about such black holes and find that the odd-parity tensor perturbations behave in a way indistinguishable to GR. On the other hand, the effective metric for the (even-parity) scalar perturbations is singular, indicating that those exact black hole solutions are infinitely strongly coupled and cannot be trusted within the regime of validity of the DHOST effective field theory. We show how this strong coupling result is generalizable to a whole class of solutions with arbitrary manifolds both for DHOST and Horndeski.
}

\maketitle

\section{Introduction}

The discovery of the cosmic accelerated expansion has motivated numerous studies on modifications of gravity in the infrared. The uniqueness of General Relativity (GR) implies that any model of model gravity necessarily need to involve additional degrees of freedom or a breaking of locality or Lorentz invariance. Including a scalar field and exploring scalar-tensor theories is therefore one of the most natural and minimalistic way to gravity beyond GR. Usually, scalar-tensor theories are considered as effective theories of more fundamental theories, and provide a general framework to explain the observed cosmic acceleration phenomenologically. Under this consideration, many attempts have been made to construct the most generic consistent theory that propagates only one scalar degree of freedom while interacting with gravity. For example, Horndeski theories \cite{Horndeski:1974wa,Deffayet:2011gz} are constructed as the most general scalar-tensor theory in four-dimensional spacetime yielding only second order field equations. In theories with multiple fields, higher-order field equations can still propagate a single degree of freedom if their Lagrangian is degenerate \cite{deRham:2011rn,deRham:2011qq}, and this argument was used to further extend scalar-tensor theories to a more general class of degenerate higher-order scalar-tensor (DHOST) theories in \cite{Langlois:2015cwa},  (see \cite{Zumalacarregui:2013pma, Motohashi:2014opa,BenAchour:2016fzp,Achour:2016rkg,Crisostomi:2016czh,deRham:2016wji,Motohashi:2016ftl} for related discussions, and \cite{Langlois:2018dxi} for a review).

On the other hand, the direct detections of gravitational waves (GWs) from binary black hole and neutron star mergers made significant effects on our understanding of gravity. Particularly, the observation of GW170817 \cite{TheLIGOScientific:2017qsa} together with its optical counterpart GRB170817A \cite{Goldstein:2017mmi} constraints the speed difference between GWs and light (propagating on a cosmological background) down to $10^{-15}$ \cite{Monitor:2017mdv}, which drastically restricts the viable candidates of scalar tensor theories \cite{Creminelli:2017sry, Sakstein:2017xjx, Ezquiaga:2017ekz, Baker:2017hug, Langlois:2017dyl}, provided that such scalar-tensor theories are still valid up to the LIGO frequency \cite{deRham:2018red}.

The development of the GW astronomy also stimulates the studies on black hole solutions in scalar-tensor theories, among which black holes with nontrivial scalar profile are of particular interests. No-hair theorem has been proved for the shift-symmetric Horndeski theory \cite{Hui:2012qt} and for the shift-symmetric Gleyzes-Langlois-Piazza-Vernizzi (GLPV) theory \cite{Babichev:2017guv}, which is a subclass of DHOST theories. Such theorems state that if the coupling functions of the theories are regular, the static, spherically-symmetric, and asymptotically flat black hole solutions with static scalar field must have the Schwarzschild metric and the constant scalar field. Hairy black holes are allowed if some of the conditions are violated. For example, by violating the regularity condition, there are asymptotically flat hairy black holes in the shift-symmetric Horndeski theory \cite{Sotiriou:2013qea,Sotiriou:2014pfa} and in the shift-symmetric GLPV theory \cite{Babichev:2017guv}. However, these solutions usually present metrics different from GR black holes. Another example is the solution found in the shift-symmetric GLPV theories that do not have the canonical kinetic term \cite{Babichev:2017guv}, which also circumvents the no-hair theorem. Also see \cite{Antoniou:2017acq,Antoniou:2017hxj,Bakopoulos:2018nui} for hairy black holes in Einstein-Scalar-Gauss-Bonnet theories.

Hairy black hole solutions can also be found if one allows for time-dependence\footnote{This point is also related to the existence of black hole solutions \cite{Berezhiani:2011mt,Rosen:2017dvn} in other models of modified gravity such as massive gravity \cite{deRham:2010kj}.}. For instance, within the context of shift-symmetric DHOST theories, a branch of hairy black holes is constructed by considering a linearly time dependent scalar field profile $\varphi(t, r) = q\, t + \psi (r)$ and a constant kinetic term $X = \partial_a \varphi\, \partial^a \varphi$. The linearly time dependent part of the scalar field can be thought as the background field that is responsible for the cosmic acceleration. Such solutions are investigated in \cite{Babichev:2013cya,Charmousis:2014zaa,Babichev:2012re,Babichev:2017guv, Kobayashi:2014eva,Babichev:2016kdt,Motohashi:2018wdq,Minamitsuji:2018vuw,Takahashi:2019oxz}. Moreover, these solutions can be ``stealth'', which means that the non-trivial scalar hair does not gravitate at the background level. Stealth black hole solutions were initially introduced in \cite{AyonBeato:2004ig}. Recently, the stealth linearly-time dependent solutions have been studied in the context of quadratic DHOST theories \cite{BenAchour:2018dap,Motohashi:2019sen}. Taking a bottom up approach, these studies identify the theories that possess stealth Schwarzschild-de Sitter (SdS) black hole solutions.

Motivated by these wealth of fruitful solutions, the stability of physical implications of those solutions were rapidly explored. If the scalar field couples directly to matter sources, one would expect binary systems to radiate scalar gravitational waves which would typically be in tension with observations unless a Vainshtein mechanism or other type of screening  is implemented \cite{Silvestri:2011ch,deRham:2012fw,deRham:2012fg,Dar:2018dra,Sagunski:2017nzb,Huang:2018pbu}. Moreover in DHOST theories, even when the scalar mode may not {\it a priori} directly couple to external matter sources, since the physical propagating degrees of freedom are mixed between the metric and the covariant scalar and implicit matter coupling typically appears in such theories which can then also lead to gravitational scalar radiation.

As for the stability of such solutions, for shift-symmetric Horndeski theories it was initially argued in \cite{Ogawa:2015pea} that they could be unstable against odd-parity perturbation, however this statement was more recently revisited in  \cite{Babichev:2017lmw, Babichev:2018uiw}. For specific time-slicings, it was argued in \cite{Ogawa:2015pea} that the positivity of either the kinetic or radial gradient energy would be violated in the vicinity of the horizon. However that particular statement is gauge dependent and in \cite{Babichev:2017lmw, Babichev:2018uiw}  it was shown how there could exist a slicing for which the theory was stable everywhere. Even if the kinetic or the gradient term where to switch sign in all slicings, this would only signal the breakdown of the Horndeski effective field theory breaks down when either the kinetic or the gradient terms become sufficiently small and the predictability of the Horndeski effective field theory therefore fails before the instability can even occur (see \cite{deRham:2017aoj} for a related discussion).

In this study we will explore the stability and validity (in the EFT sense) of the exact quadratic DHOST solution found in \cite{BenAchour:2018dap, Motohashi:2019sen}. Odd perturbations have already been the subject of a very interesting analysis  \cite{Takahashi:2019oxz} (which appeared during the final stages of this work), and in the rest of this paper we shall investigate both the odd and even-parity perturbations about the stealth black hole solutions in the shift-symmetric quadratic DHOST theories. In agreement with  \cite{Takahashi:2019oxz}, we show that such solutions are stable against odd-parity perturbations. Indeed, the odd-parity perturbations are exactly the same as that of GR black holes. The even-parity perturbations however differ due to the presence of the scalar field and while the diagonalized would-be tensor modes could a priori be stable, the diagonalized scalar mode sees a singular effective metric, indicating the stealth black hole solution in the shift-symmetric quadratic DHOST theories can unfortunately not be trusted. \\

To put the current findings in perspective, we highlight that the nature of this problem is more severe than that observed in \cite{Ogawa:2015pea} for Horndeski black holes. Indeed fluctuations about the shift-symmetric Horndeski black hole solutions are well-behaved at sufficiently large distances and the theory only becomes unreliable close to the horizon (where either the kinetic or gradient term becomes small, before they would become negative). Such solutions can therefore still potentially provide relevant phenomenology away from the horizon. For the stealth black hole solutions in the shift-symmetric quadratic DHOST theories on the other hand, the effective scalar fluctuations are always everywhere and all the time degenerate and those solutions can therefore never and nowhere be trusted. These types of pathologies are very similar in nature to those observed about the exact static black solutions found in massive gravity \cite{Rosen:2017dvn}, where it was shown that solutions that perfectly mimic GR black hole solutions, actually do so by effectively suppressing the graviton mass on those backgrounds and therefore making the additions degrees of freedom present in massive gravity infinitely strongly coupled. The case of DHOST appears to be very analogous in that black hole solutions can only precisely the same as in GR if the effect of the scalar field is effectively entirely suppressed and therefore making this degree of freedom infinitely strongly coupled. Instead a perhaps more promising direction for black hole solutions in DHOST theories just like in massive gravity is the presence of a (small but nonetheless nonzero) time dependence of the metric. This time-dependence would typically be governed by the graviton mass scale in massive gravity or the dark-energy scale in DHOST theories and could therefore imply a time-evolution only visible on time scales of the order of the age of the Universe\footnote{This point is also tightly linked to the need of small yet non-vanishing amount of space-dependence for consistent massive cosmology solutions \cite{DAmico:2011eto}.}. \\

The rest of the paper is organized as follows. In Sec.~\ref{sec:BH}, we first present the shift-symmetric quadratic DHOST theories and review the linear-time dependent black hole solutions. In Sec.~\ref{sec:pert}, we derive equations of both odd and even parity perturbations and highlight the presence of an (infinitely) strongly coupled degree of freedom. The strong coupling issue is then generalized to a larger class of solutions with generic manifolds in DHOST theories including rotating black hole solutions in Sec.~\ref{sec:SC}. The implications for a class of solutions in Horndeski is also highlighted. Sec.~\ref{sec:dis} is then conclusions and outlook.

\section{Stealth Black Holes in Shift-Symmetric DHOST Theories}\label{sec:BH}

The action of the shift-symmetric DHOST theory up to quadratic order is given by \cite{Langlois:2015cwa},
\ba\label{eq:DHOST}
S_{\rm qDHOST} = \int \d^4x \sqrt{-g}\,\left[ P(X)+ Q(X) \Box\varphi + F(X) {\R} + \sum_{i=1}^{5} A_i(X) B_i \right],
\ea
where
\ba
&&B_1 =\varphi_{ab} \varphi^{ab}, \quad B_2 = \left(\Box\varphi\right)^2, \\
&&B_3 = \varphi^a \varphi^b \varphi_{ab}\,\Box\varphi, \quad B_4 = \varphi^a\varphi_{ab}\varphi^{bc}\varphi_c, \quad B_5=\left(\varphi_{ab}\varphi^{a}\varphi^{b}\right)^2,
\ea
with $\varphi_a = \p_a \varphi$, $\varphi_{ab}=D_a D_b \varphi$ and $X= \varphi_a \varphi^a=g^{ab} \varphi_a \varphi_b$.  In the wake of GW170817 \cite{Langlois:2017dyl}, the requirement that the speed of GWs should be the same as light in a cosmological background imposes the following conditions (if one were to assume that the DHOST effective field theory remained sufficiently under control at LIGO frequency scale \cite{deRham:2018red}),
\ba
&&A_1 = A_2 = 0, \quad A_4= \frac{1}{8F} \left[48 F'^2 - 8\left(F-XF'\right)A_3 - X^2 A_3^2\right], \nonumber \\
&&A_5 = \frac{1}{2F} \left(4F'+X A_3\right)A_3,
\label{eq:GWcconstraints}
\ea
where here a prime denotes the derivative with respect to the argument $X$. In addition, it also requires $A_3=0$ to prevent a rapid decay of GWs into the scalar field \cite{Creminelli:2018xsv} (see also \cite{Creminelli:2019nok}). This condition as well the condition $A_1=0$ were not imposed in the recent analysis presented in \cite{Takahashi:2019oxz}.\\

Thus the subclass of DHOST theories will be considered in this paper is
\ba\label{eq:DHOST2}
S = \int \d^4x \sqrt{-g}\,\left[ P(X)+ Q(X) \Box\varphi + F(X) {\R} +\frac{6 F'^2}{F} \varphi^2_{ab}\varphi^a \varphi^b + {\cal L}^{(\text{matter})}\left(g, \psi_i\right)\right]\,,
\ea
where we used $A_4 = 6 F'^2/F$.
Note that we have also included external matter fields $\psi_i$ that only couple to the metric $g\mn$.
Even though at this level there is no direct coupling between  $\varphi$ and the external matter sources, we will still see in what follows that the physical scalar degree of freedom in this theory does directly couple (already at tree-level) to external sources. This is due to the non-trivial mixing of the physical degrees of freedom in these types of degenerate field theories.\\

For the back hole solutions, we assume the ansatz,
\ba\label{eq:ansatz}
\d s^2 = \g_{ab} \d x^a \d x^b =- A(r) \d t^2 + \frac{1}{A(r)} \d r^2 + r^2 \d\Omega^2, \quad \text{and} \quad \bar \varphi = q t + \psi(r),
\ea
Note that  $A(r)$ has nothing to do with the $A_i(X)$ previously introduced in \eqref{eq:DHOST}.
We further require $X$ to be a constant, $X = X_0$, which implies \cite{Babichev:2013cya}
\ba\label{eq:profile}
\frac{d\psi}{dr} = \pm \frac{1}{A}\sqrt{q^2 + X_0 A}\,.
\ea
Under the ingoing Eddington-Finkelstein coordinates $(v, r)$ defined by $\d v = \d t + \d r/A$, it can be shown that $\varphi \simeq q v$ at the vicinity of the future event horizon \cite{Babichev:2013cya, Motohashi:2019sen}. Therefore, we will pick the branch with the ``$+$" sign, in which case $\varphi$ is regular at the future event horizon. In this paper, we will mostly focus on SdS black holes, i.e.
\ba\label{eq:A}
A(r) = 1- \frac{2M}{r} - \Lambda r^2.
\ea
The theories also allow Schwarzschild black holes with linear time-dependence, which can be obtained by sending $\Lambda \rightarrow 0$ \cite{Motohashi:2019sen}, and our analysis applies as well. Following the analysis presented in \cite{Motohashi:2019sen, BenAchour:2018dap}, we can see that the ansatz (\ref{eq:ansatz}) solves the equation of motions in two following cases:
\ba
&&\text{Case 1:} \quad X_0=-q^2, \quad P +6\Lambda F = 2 P' + 24 \Lambda F' - 9 \Lambda q^2 A_3 = Q'=0\, , \label{case1} \\
&&\text{Case 2:} \quad X_0 \neq -q^2, \quad P +6\Lambda F =  P' + 12 \Lambda F'  = Q' = A_3 =0\, ,\label{case2}
\ea
where all terms are evaluated at $X=X_0$. For theories with $A_3 = 0$, i.e. theories have no significant GWs to scalar filed decay, the conditions of having black holes in those two cases are degenerate.

\section{Black Hole Perturbations}\label{sec:pert}

We first start with the covariant equations of motion. Varying the action (\ref{eq:DHOST2}) with respect to the inverse metric $g^{ab}$, we obtain the modified Einstein equation,
\ba\label{eq:Eab0}
{\cal E}_{ab} &=& F G_{ab} + F'\left(\R \varphi_a \varphi_b - D_a D_b X +\Box X g_{ab}\right) -F''\left(X_aX_b-\left(\partial X\right)^2 g_{ab}\right)  \\
&+& P' \varphi_a\varphi_b -\frac12 P g_{ab}- \frac12 Q'\left(X_a\varphi_b + \varphi_a X_b - X_c\varphi^c \,g_{ab} - 2\,\Box \varphi \, \varphi_a \varphi_b\right) \nonumber \\
&-&\frac18 A_4\left(4\Box X \varphi_a \varphi_b-2X_aX_b+\left(\partial X\right)^2 g_{ab}\right) -\frac14 A_4'\left(\partial X\right)^2 \varphi_a \varphi_b - T_{ab} = 0, \nonumber
\ea
where we define $X_a \equiv \p_a X$. Eq.~(\ref{eq:Eab0}) involves terms with three derivatives acting on $\varphi$. Here $T_{ab}$ is the (conserved) stress-energy tensor associated with the matter fields $\psi_i$. Conservation of $T_{ab}$, i.e. $\D_a T^a{}_b=0$ sets the equation of motion for the matter fields $\psi_i$. Varying with respect to $\varphi$ yields the equation of motion for the scalar field
\ba\label{eq:Ephi0}
{\cal E}_{\varphi} = D_a\left[Q' X^a-2\left(F' \R + P' + Q' \Box \varphi - \frac12 A_4 \Box X - \frac14 A_4' \left(\partial X\right)^2 \right) \varphi^a \right] = 0.
\ea
To perturb the equations about the black hole solution, we write $g_{ab} = \g_{ab}+h_{ab}$, $\varphi=\bar{\varphi}+ \delta\varphi$, and $T_{ab} = 0 + \delta T_{ab}$, where a subscript ${}_0$ or a bar refers to the background\footnote{To be technically correct, the background expression for $T_{ab}$ does not actually vanish everywhere, it is a delta function at the origin, scaling with $M$ and corresponding to the physical source of the back hole. However away from the origin the background part of the stress-energy tensor vanishes for the black hole situation considered here and is therefore irrelevant for the rest of this study.}. For convenience, we also keep $X= X_0 + \delta X$ with $\delta X = 2\bar{\varphi}^a \delta\varphi_a - \bar{\varphi}^a\bar{\varphi}^b\, h_{ab}$ where indices are raised and lowered with respect to the background SdS metric $\g_{ab}$. Working to first order in perturbations about the background, we then have the perturbed equations
\ba\label{eq:Eab1}
{\cal E}_{ab}^{(1)} = &&F_0\, \delta G_{ab} + 3 \Lambda F_0\, h_{ab}+ F_0' \,\bar{\varphi}_a \bar{\varphi}_b\, \delta \R - \delta T_{ab} \nonumber \\
&& +\left[3\Lambda F_0' \,\bar{g}_{ab} + \left(F_0''\R_0 + P_0'' + Q_0'' \Box \bar{\varphi} \right)\bar{\varphi}_a \bar{\varphi}_b\right] \delta X \nonumber \\
&& - F_0' \,\delta X_{ab} + \left(F_0' \, \bar{g}_{ab} - \frac{3 F_0'^2}{F_0}  \bar{\varphi}_a \bar{\varphi}_b\right) \Box \delta X = 0,
\ea
and
\ba\label{eq:Ephi1}
{\cal E}_{\varphi}^{(1)} = \D_{a}\left\{\bar{\varphi}^a\left[-2F_0'\,\delta \R + \frac{6F_0'^2}{F_0} \Box\delta X -2  \left(F_0''\R_0 + P_0'' + Q_0'' \Box \bar{\varphi} \right)\delta X\right]\right\} = 0,
\ea
where we have used conditions (\ref{case1}) or (\ref{case2}) and $\R_0 = 12 \Lambda$, and have defined $\delta X_{ab}=D_a D_b \delta X$. To make further progress and properly separate out the relevant degrees of freedom, we consider the following linear combinations of the equations of motion:
\ba\label{eq:combin}
&&(F_0-F_0'X_0)\,{\cal E}_{\varphi}^{(1)} -2 F_0 \, \D_a \! \left({\cal E}^{(1)}\  \varphi^a \right) = 0,  \\
\text{and} \quad && {\cal E}_{ab}^{(1)} - \frac{F_0'}{F_0'X_0 - F_0}\bar{\varphi}_a\bar{\varphi}_b\, {\cal E}^{(1)} + \frac12 \frac{F_0}{F_0'X_0 - F_0} \bar{g}_{ab} \,{\cal E}^{(1)} = 0\,,
\ea
where we used the notation ${\cal E}^{(1)}\equiv {\cal E}^{(1)} {}^c_{\, c}= \bar{g}^{ab} {\cal E}^{(1)}_{ab}$, leading to
\ba\label{eq:Ephi2}
\D_a \left[ \bar{\varphi}^a \( -2F_0  \left(F_0''\R_0 + P_0'' + Q_0'' \Box \bar{\varphi} + 2 \bar{A_4} \right) \delta X + 4 F_0'\, \delta T \) \right] = 0,
\ea
and
\ba\label{eq:Eab2}
F_0\left(\delta \R_{ab} - 3\Lambda h_{ab}\right) = F_0 S_{ab}
\ea
respectively. Here we have defined the source tensor $S_{ab}$ as
\ba\label{eq:Sab}
F_0\, S_{ab} \equiv&& \delta T_{ab} + \frac{1}{F_0'X_0-F_0}\left(\frac{1}{2}F_0 \bar{g}_{ab} - F_0' \bar{\varphi} _a\bar{\varphi}_b\right)\delta T + \frac{1}{2}F_0' \bar{g}_{ab} \Box \delta X + F_0' \delta X_{ab} \nonumber \\
-&& \left[3\Lambda F_0' + \frac{6\Lambda F_0 F_0'}{F_0'X_0-F_0} + \frac12 \frac{F_0 X_0}{F_0'X_0-F_0}\left(F_0''\R_0 + P_0'' + Q_0'' \Box \bar{\varphi}\right)  \right] \bar{g}_{ab}\delta X \nonumber \\
+&&\left[ \frac{12\Lambda F_0'^2}{F_0'X_0-F_0} + \frac{ F_0 }{F_0'X_0-F_0}  \left(F_0''\R_0 + P_0'' + Q_0'' \Box \bar{\varphi}\right)\right] \bar{\varphi}_a\bar{\varphi}_b \delta X .
\ea
In the following,  we shall decompose the perturbations based on their behaviours under parity transformations $\left(\theta,\, \phi \right) \rightarrow \left(\pi-\theta, \, \phi+\pi \right)$. This decomposition allows us to consider odd perturbation and even perturbation separately.

\subsection{Odd sector}\label{sec:oddpert}

 The odd-parity perturbation of the metric can be written as \cite{PhysRev.108.1063, Zerilli:1970se, Motohashi:2011pw}
 \ba\label{eq:habodd}
h_{ab}^{\text{odd}}= \begin{pmatrix}
0 &0  &- {\rm h}_0 \csc \theta \, Y_{\ell m,\phi} & {\rm h}_0 \sin \theta \, Y_{\ell m,\theta}  \\
0 & 0  & - {\rm h}_1 \csc \theta \, Y_{\ell m,\phi}& {\rm h}_1 \sin \theta \, Y_{\ell m,\theta} \\
- {\rm h}_0 \csc \theta \, Y_{\ell m,\phi} & - {\rm h}_1 \csc \theta \, Y_{\ell m,\phi}& \frac{1}{2} {\rm h}_2 \csc \theta \,\mathcal{X} &  -\frac{1}{2} {\rm h}_2 \sin \theta \, \mathcal{W}\\
 {\rm h}_0 \sin \theta \, Y_{\ell m,\theta}&  {\rm h}_1 \sin \theta \, Y_{\ell m,\theta}&  -\frac{1}{2} {\rm h}_2 \sin \theta \,\mathcal{W} &  -\frac{1}{2} {\rm h}_2 \sin \theta \, \mathcal{X}
\end{pmatrix} , \label{odd metric perturbation}
\ea
where ${\rm h}_0$, ${\rm h}_1$, ${\rm h}_2$ are functions of $(t, r)$, $Y_{\ell m}$ are the spherical Harmonics, a comma denotes the partial derivate, and
\ba
\begin{aligned}
\mathcal{X} &= 2(\partial _ \theta \partial _ \phi - \cot \theta \partial _ \phi  )Y_{\ell m}\,,
\\
\mathcal{W} &= (\partial _ \theta \partial _ \theta - \cot \theta \partial _ \theta  - \csc ^2 \theta  \partial _ \phi\partial _ \phi  )Y_{\ell m}\,.
\end{aligned}
\ea
Note that perturbation of the scalar field $\delta\varphi$ is even under the parity transformation, hence does not couple with $h_{ab}^{\text{odd}}$ and can be omitted in the odd perturbation equations. Moreover, explicate calculation shows that $h^{\text{odd}}_{ab}\varphi^a\varphi^b=0$ and hence $\delta X$ vanishes in the odd sector. Together with the fact that $\delta \R^{\text{odd}} =0$,  Eq.~(\ref{eq:Eab1}) simplifies to
\ba
F_0\left(\delta G_{ab} + 3\Lambda h_{ab}\right) - \delta T_{ab} = 0\,,
\ea
where $F_0$ plays the effective role of the Planck scale and we can therefore conclude that the odd parity perturbation of the stealth SdS black holes considered here is identical to that in GR.

\paragraph{Relaxing the assumptions:} The previous result relied on the assumption $A_3=0$ (so as to prevent GWs from decaying into dark energy), however we may wonder what the effects would be if some of those assumptions were relaxed.
Actually, we find that even in the case where $A_3 \neq 0$ and hence $A_5 \neq 0$, the odd sector of GWs still behave identically as in GR. We can see this by perturbing the Lagrangian to quadratic order and using the fact that $\bar{B}_3 = \bar{B}_5 = \delta^{(2)}B_5 = 0$, then we can see that the presence of $A_3$ and $A_5$ leads to two extra terms, $9 q^2 \Lambda \bar{A}_3 \bar{\varphi}^a \bar{\varphi}^b \g^{cd} h^{\text{odd}}_{ac}h^{\text{odd}}_{bd}$ and $\bar{A}_3\, \delta^{(2)}B_3$, which eventually cancel each other given the background solution\footnote{While this work was in progress, the interesting analysis of \cite{Takahashi:2019oxz} appeared on the arXiv, also discussing the odd perturbation of stealth black holes but with further relaxing the assumptions to $A_1 \neq 0$. Our results agree in the case of $A_1 = 0$.}.

\subsection{Even sector}\label{sec:evenpert}

Next we turn to  the even parity perturbations. The metric perturbations can be written as \cite{PhysRev.108.1063,Zerilli:1970se,Motohashi:2011pw}
\begin{equation}
h_{ab}^{\text{even}} = \begin{pmatrix}
A H_0 & \hspace{0.2cm}& H_1  & \hspace{0.2cm}& \mathcal{H}_0 \partial_\theta & \hspace{0.2cm}&  \mathcal{H}_0 \partial_\phi \\
H_1 & \hspace{0.2cm}& H_2/A  & \hspace{0.2cm}& \mathcal{H}_1 \partial_\theta & \hspace{0.2cm}& \mathcal{H}_1 \partial_\phi \\
\mathcal{H}_0 \partial_\theta & \hspace{0.2cm}& \mathcal{H}_1 \partial_\theta & \hspace{0.2cm}& \mathcal{K} + \G \nabla_\theta \nabla_\theta & \hspace{0.2cm}& \G \nabla_\theta \nabla_\phi\\
\mathcal{H}_0 \partial_\phi & \hspace{0.2cm}& \mathcal{H}_1 \partial_\phi & \hspace{0.2cm}& \G \nabla_\phi \nabla_\theta & \hspace{0.2cm}& \sin^2\theta \mathcal{K}+ \G \nabla_\phi \nabla_\phi
\end{pmatrix} Y_{\ell m} \, , \label{eq:evenhab}
\end{equation}
where again $H_0, H_1, H_2, \mathcal{H}_0, \mathcal{H}_1, \mathcal{K}$ and $\G$ are functions of $(t,r)$, and $\nabla_{\theta,\, \phi}$ are covariant derivatives on the $2$--sphere of radius one. We now also have the scalar perturbation
\ba
\delta \varphi = \Phi\left(t,r\right) Y_{\ell m}.
\ea
In the following, we will sketch how to solve the even perturbation. An observation is that, given some initial conditions\footnote{Note that not all components in $h_{ab}^{\text{even}}$ are independent as most of them are related constraint equations as we shall see later. To solve for the system, we only need to set initial conditions for $\chi$, $\dot{\chi}$, $\delta X$ and $\Phi$.} for $\delta \varphi$ and $h_{ab}^{\text{even}}$, one can directly solve for $\delta X$ as a whole from Eq.~\eqref{eq:Ephi2} (for a specific matter source distribution set by $\delta T_{ab}$). The expression for $\delta X$ can then be plugged into the effective source term $S_{ab}$ defined in Eq.~\eqref{eq:Eab2} and this can then be used to solve for the remaining even-parity effective tensor mode in a very similar way as in GR.  To see this work in practise it is convenient to set a gauge and we do so differently depending on whether we are dealing with the monopole, dipole or higher multipoles.

\subsubsection{Higher multipoles}

For multipoles with $\ell \ge 2$, we may fix the gauge by setting $\G = \K = \scH_0 = 0$ and refer to appendix \ref{app:diff} to see how coordinate transformations affect the even sector and check that this gauge can be chosen.
The gauge fixing for monopole and dipole are different and will be treated separately below.

Having fixed  $\G = \K = \scH_0 = 0$  we can then derive explicitly the master equation for one of the propagating  degrees of freedom (effectively the even-parity tensor) by replacing $H_2$ with a new variable $\chi$ defined through,
\ba\label{H2}
H_2 = \frac{\ell (\ell+1)}{r}\scH_1 - \frac{1}{A r}\chi.
\ea
For convenience, we denote
\ba
{\cal E}_{ab}^{L}\equiv \delta \R_{ab} - 3\Lambda h_{ab},
\ea
so that Eq.~\eqref{eq:Eab2} can be written as ${\cal E}_{ab}^{L} = S_{ab}$. The equation for $\chi$ can be obtained by considering the following combination,
\ba\label{eq:chiL}
\ddot{\chi} - A^2 \chi_{,rr} + f_1 \chi_{,r} + f_2 \chi &\equiv & c_1 \frac{{\cal E}_{tt}^{L}}{Y} + c_2 \frac{{\cal E}_{rr}^{L}}{Y} + c_3 \left(\frac{{\cal E}_{\theta\theta}^{L}}{Y}+\frac{{\cal E}_{\phi\phi}^{L}}{\sin^2\theta Y} \right) +c_4 \frac{{\cal E}_{r\theta}^{L}}{Y_{,\theta}}  \\
&+& c_5 \frac{{\cal E}_{\theta\phi}^{L}}{\cot \theta Y_{,\phi}- Y_{,\theta\phi}} + d_1 \frac{{{\cal E}_{tt}^{L}}_{,r}}{Y} + d_2 \frac{{{\cal E}_{rr}^{L}}_{,r}}{Y} + d_3\left(\frac{{\cal E}_{\theta\theta}^{L}}{Y}+\frac{{\cal E}_{\phi\phi}^{L}}{\sin^2\theta Y} \right)_{,r}+d_4  \frac{\dot{{\cal E}_{tr}^{L}}}{Y}\,, \nonumber
\ea
where commas denote partial derivatives and the coefficients $c_i$, $d_i$ and $f_i$ are given in Appendix \ref{app}. Then the equation of motion for $\chi$ can be written as
\ba\label{eq:chi}
\ddot{\chi} - A^2 \chi_{,rr}+ f_1 \chi_{,r} + f_2 \chi = s(t,r)
\ea
where $s(t,r)$ depends $\delta X$ and $\delta T_{ab}$, and is given by the same combination as the right hand side of Eq.~\eqref{eq:chiL} with ${\cal E}_{ab}^{L}$ replaced by $S_{ab}$. Using the following relations
\ba
&&S_{tt},\, S_{rr},\, S_{tr} \propto Y \\
&&F_0S_{r\theta} = F_0' \delta X_{r\theta} + \delta T_{r\theta} \propto Y_{,\theta}\\
&&F_0S_{\theta\phi} = F_0' \delta X_{\theta\phi} + \delta T_{\theta\phi} \propto Y_{,\theta\phi}-\cot\theta Y_{,\phi}\\
&&F_0\left(S_{\theta\theta}+\frac{S_{\phi\phi}}{\sin^2\theta}\right) \propto Y, \label{eq:propang}
\ea
the angular dependence fully drops out from the right hand side of  Eq.~\eqref{eq:chi}.

In particular the last relation \eqref{eq:propang} can be seen as follows. For $a, b = \theta, \phi$, terms in $S_{ab}$ proportional to $\bar{\varphi}_{a}\bar{\varphi}_b$ vanish, terms proportional to $\bar{g}_{ab}$ lead to contribution proportional to $r^2 Y$, and terms involving covariant derivatives form the Laplacian operator in the $2$--sphere and therefore lead to contribution proportional to $\ell(\ell+1) Y$.

We can now (in principle) solve \eqref{eq:chi} explicitly for  $\chi$, and infer  the other components in the metric perturbations by considering the following constraint equations:
\ba
&&\frac{r^2}{A} \frac{{\cal E}_{tt}^{L}}{Y} + A r^2  \frac{{\cal E}_{rr}^{L}}{Y} +\left(\frac{{\cal E}_{\theta\theta}^{L}}{Y}+\frac{{\cal E}_{\phi\phi}^{L}}{\sin^2\theta Y} \right) = -2 \chi_{,r} - \frac{J}{rA}\chi -\frac{J(3A+3\Lambda r^2-J-1)}{r}\scH_1,\\
&&\frac{2r^2}{J} \frac{{\cal E}_{tr}^{L}}{Y}  = \dot{\scH_1}- \frac{2}{JA}\dot{\chi} + H_1,\\
&&\frac{2{\cal E}_{\theta\phi}^{L}}{\cot \theta Y_{,\phi}- Y_{,\theta\phi}}= -2A \scH_{1,r}+ \left(\frac{J}{2}-2A_{,r}\right)\scH_1 - \frac{1}{rA}\chi - H_0,
\ea
where $J=\ell(\ell+1)$. Again, the above constraint equations are accompanied with some ``source" terms on the their right hand side, which are given by a same combination with ${\cal E}_{ab}^{L}$ replaced by $S_{ab}$, hence proving the constraints for $H_0$, $\scH_1$, $H_1$ and $H_2$. With this in mind we can then eventually solve the remaining dynamical degree of freedom (namely the scalar degree of freedom) $\Phi$ by using
\ba
\delta X = \left[\frac{2q^2\sqrt{1-A}}{A}H_1 + \frac{q^2\left(A-1\right)}{A}H_2 - \frac{q^2}{A}H_0 + 2 q\sqrt{1-A}\Phi_{,r} - \frac{2q}{A}\dot{\Phi} \right] Y.
\ea
The above analysis shows that, for multipoles with $\ell \ge 2$, there are two propagating degrees of freedom in the even sector. With a trivial choice of $P$, $Q$ and $F$, we can get back to GR, in which case $\chi$ becomes the usual propagating degree of freedom in the even-parity sector. Therefore, we may think of the dynamical equations for the two degrees of freedom in the even-parity sector as being Eqs.~\eqref{eq:Ephi2} and \eqref{eq:chi}. We shall comment in this in what follows but first we look at the monopole and dipole.

\subsubsection{Monopole}

In the case of monopole, the contributions from $\scH_0$, $\scH_1$ and $\G$ vanish identically. We therefore instead set the gauge $\K = H_1 =0$ (see appendix~\ref{app:diff} for confirmation that such a gauge can be fixed for the monopole). Then we  find the following two constraints equations
\ba
&&\frac{r^2}{A} \frac{{\cal E}_{tt}^{L}}{Y} + A r^2  \frac{{\cal E}_{rr}^{L}}{Y} +\left(\frac{{\cal E}_{\theta\theta}^{L}}{Y}+\frac{{\cal E}_{\phi\phi}^{L}}{\sin^2\theta Y} \right) = 2 A r H_{2,r}+ \left(2-6\Lambda r^2\right)H_2\\
&&\frac{r}{A^2}\frac{{\cal E}_{tt}^{L}}{Y}  + r \frac{{\cal E}_{rr}^{L}}{Y} = H_{2,r}-H_{0,r},
\ea
which confirms the fact that there are no monopole tensor modes and the relevant dynamics of the physical scalae monopole is given once again by Eq.~\eqref{eq:Ephi2}.
\subsubsection{Dipole}

In the case of dipole, $h_{ab}^{\text{even}}$ only depends on $\K$ and $\G$ through the particular combination $\K-\G$, and thus we can set $\K = \G = \scH_0 =H_2=0$ by fixing gauge (see again appendix~\ref{app:diff} for confirmation that such a gauge can be fixed for the dipole). The other component can be solved by the constraint equations below:
\ba
&&\frac{r^2}{A} \frac{{\cal E}_{tt}^{L}}{Y} + A r^2  \frac{{\cal E}_{rr}^{L}}{Y} +\left(\frac{{\cal E}_{\theta\theta}^{L}}{Y}+\frac{{\cal E}_{\phi\phi}^{L}}{\sin^2\theta Y} \right) = -4A \scH_{1,r} - \frac{2(1+A-3\Lambda r^2)}{r}\scH_1\\
&&r^2 \frac{{\cal E}_{tr}^{L}}{Y}  = - \dot{\scH_1} + H_1\\
&&\frac{2{\cal E}_{\theta\phi}^{L}}{\cot \theta Y_{,\phi}- Y_{,\theta\phi}}= -2A\scH_{1,r}-2A_{,r}\scH_1-H_0,
\ea
which also confirms the fact that there are no dipole tensor modes and the relevant dynamics of the physical scalar dipole is given also once again by Eq.~\eqref{eq:Ephi2}.

\subsubsection{Dynamics of the scalar mode}

Whether we were dealing with the monopole, the dipole or the higher multipoles, we have shown that the relevant dynamics for the physical scalar mode is governed solely by Eq.~\eqref{eq:Ephi2}. With this in mind, we shall therefore focus on that equation more closely and  instead of any of the gauge choices we used previously,  we shall now set a  gauge so that $\bar{\varphi}^a\bar{\varphi}^b h_{ab}^{\text{even}}=0$ irrespectively of which multipole we are dealing with. We emphasize that this is only for convenience but none of the results depends on that precise gauge choice. In this case, it is easy to see that  Eq.~\eqref{eq:Ephi2} becomes solely an equation for $\delta\varphi$ of the form
\ba
\label{eq:singeq}
\bar{\varphi}^a  \bar{\varphi}^b \D_a\D_b\delta\varphi + \left(\bar{\varphi}_a  \bar{\varphi}^{ab}+ \Box\bar{\varphi}\bar{\varphi}^b + \frac{\bar{\varphi}^a \D_a \bar{\Omega}(r)}{\bar{\Omega}(r)}\bar{\varphi}^b\right)\p_b \delta\varphi = \frac{1}{\bar{\Omega}(r)} D_a \left(\frac{F_0'}{F_0}\bar{\varphi}^a \delta T\right),
\ea
where $\bar{\Omega}(r) = 12\Lambda F_0'' + P_0'' +  Q_0''\Box \bar{\varphi} + 12 \Lambda F_0'^2/F_0$. Given the background solution, we see that the scalar fluctuation $\delta \varphi$ sees a singular effective metric $g_{\rm eff}^{ab}\sim \bar{\varphi}^a \bar{\varphi}^b$, which only ever has one non-vanishing eigenvalue. This implies that the physical (diagonalized) scalar fluctuations living on this exact black hole solution would are thus infinitely strongly coupled and the background solution cannot be trusted.

\section{Strong Coupling Issues for Generic Manifolds}
\label{sec:SC}

\paragraph{Quadratic DHOST:}   Before concluding, it is worth pointing out that the strong coupling results derived in this manuscript hold beyond the SdS metric considered here and are actually generalizable to much more generic manifolds\footnote{We wish to thank Hayato Motohashi for very useful discussion on this point.} and scenarios so long as $X$ is constant on the background manifold, $X=X_0=$const.\\

Indeed consider the full quadratic DHOST theory \eqref{eq:DHOST} with generic shift-symmetric functions $P,F,Q$ and $A_i(X)$, then we can show that any background solution (denoted by the subscript $_0$) that satisfies the following properties {\it on that particular background solution} suffers from infinitely strong coupling and cannot be trusted
\ba
\label{eq:SC1}
\left\{
\begin{array}{l}
X_0={\rm const}\,,\\[5pt]
A_1(X_0)=A_2(X_0)=A'_1(X_0)=A'_2(X_0)=0\\[5pt]
A_3(X_0)=A_5(X_0)=Q'(X_0)=0, \\[5pt]
A_4(X_0)=6\frac{F'(X_0)^2}{F(X_0)}\,, \\[5pt]
P'(X_0)=-R_0 F'(X_0)\,.
\end{array}\right.
\ea
In particular this implies that the rotating black hole solutions found in \cite{Charmousis:2019vnf} suffer from the same issue, and fluctuations of the scalar degree of freedom about the rotating black hole found in  \cite{Charmousis:2019vnf} with finite stealth hair  is infinitely strongly coupled (apart in the spacial case of spherical symmetry where the constraint $A_3(X_0)=0$ is relaxed).

Note that those conditions do not impose to be dealing with theories where for instance $A_3, A_5$ and $Q'$ vanish identically, and strong coupling would still be an issue even if say $A_3'(X_0)\ne 0 $ or $A_5'(X_0)\ne 0$ or $A_1''(X_0)\ne 0$.  To avoid strong coupling at least one of the constraints  in \eqref{eq:SC1} should be violated but note also that the conditions \eqref{eq:SC1} are not the unique conditions under which the issue may arise, and violating one or several of the conditions in \eqref{eq:SC1} does not necessarily ensure the absence of strong coupling issue.  It is possible that strong coupling occurs on particular solutions even if the previous conditions are not satisfied, or that strong coupling arises instead for the tensor degree of freedom \cite{deRham:2016wji}.

\paragraph{Further relaxing the assumptions:}  One may raise the question of what would occur if for instance $A_3(X_0)$ did not vanish precisely but was simply taken to be (extremely) small so as to prevent too much GW decay into dark energy on a particular solution of interest (of course if the EFT is not valid on those scales, the constraint on $A_3$ could potentially be relaxed further). If for instance $A_3(X_0)$ was considered to be small but non-vanishing,  then the effective metric of the scalar degree of freedom about that solution would could include four non-vanishing eigenvalues but the magnitude of those would be governed by the (extremely) small scale present in $A_3(X_0)$ and would also indicate strong coupling issues (low cutoff). For instance if one were to consider perturbations about a spherically symmetric configuration,  this would imply that the higher multipoles would not be suppressed as compared to lower multipoles. Second since $\delta \varphi$ does couple to the trace of external matter fields in generic DHOST theories as can be seen from the right hand side of \eqref{eq:singeq}, any small test particle would lead to arbitrarily large  emission of scalar waves\footnote{Unless $F_0'=0$ in which case \eqref{case1} or \eqref{case2}  would also imply $Q_0'=P_0'=0$ and we would have $A_4(X_0)=0$, then on that background $P_0$ would effectively play the role of a cosmological constant, the term proportional to $Q_0$ would be a total derivative and we would effectively just be dealing with GR and a scalar field minimally coupled to gravity.}.

\paragraph{Horndeski:}  This also applies to any Horndeski theories \cite{Horndeski:1974wa} that satisfies an equivalent set of conditions, independently of how symmetric (or not) the background manifold is. Consider a shift-symmetric Horndeski theory of the form
\ba
\label{eq:Horndeski}
S_{\rm Horndeski}&=& \int \d^4 x \sqrt{-g}\Bigg[
K(X)-G_3(X)\Box \varphi
-2G_4(X)R+G_{4}'(X)\((\Box \varphi)^2-\varphi_{ab}^2\)\\
&&+G_5(X)G_{ab} \varphi^a \varphi^b+\frac 13 G_{5}'(X)
\((\Box \varphi)^2-3 \Box \varphi \varphi_{ab}^2+2 \varphi_{ab}^3\)
\Bigg]\,,\notag
\ea
with minimal coupling to external sources.
Then for this theory, any  solution  on  which $X_0$ is constant and for which $G_{3}'(X_0)=G_{4}''(X_0)=G_{5}'(X_0)=G_{5}''(X_0)=K'(X_0)=0$, the scalar propagating degree of freedom about this would be solution is infinitely strongly coupled and the existence of such a solution could not be trusted.  This result is independent of any details of the manifold considered and the symmetry of the solution.
This is in addition to potential strong coupling issue that may occur for the tensor modes. \\

In particular, we may point out that for appropriate choices of functions $A_i$, the DHOST theory considered in \eqref{eq:DHOST} reduces to a special case of Horndeski \cite{Horndeski:1974wa}, for which the static black hole solutions and their stability were explored in \cite{Kobayashi:2014wsa} and the strong coupling results remain valid in that particular subclass of Horndeski.

Indeed, following the analysis performed in \cite{Kobayashi:2014wsa} (which applied for static solutions $q=0$), with $G_{3,X}(X_0)=K_{,X}(X_0)=0$, $G_4$ to be constant  and imposing $G_5$ to vanish we find that the effective metric for the dynamical even-degrees of freedom is singular, see appendix~\ref{app:HorndeskiBH}, confirming a strong coupling issue for that particular limit of the Horndeski black Hole solutions. In that case, this strong coupling issue appears to be closely linked to the requirement that $X$ be a constant at the background level.

\section{Outlooks}\label{sec:dis}

In this paper, we investigated the perturbation of linearly time-dependent stealth SdS black holes in shift-symmetric quadratic DHOST theories. We focus on the subclass of DHOST theories described by action (\ref{eq:DHOST2}), i.e. those theories that (1) predict unitary GW speed, and (2) have no significant decay of GWs into the scalar fluctuations. The linearly time-dependent stealth SdS black holes exist if the functions in action (\ref{eq:DHOST2}) satisfy conditions (\ref{case1}) or (\ref{case2}). As usual, we decomposed the perturbation based on their parity and derived the perturbation equations respectively.\\

By deriving the perturbation equations, we find that the odd-parity perturbation is the same as that of GR black holes. Actually, this is the case even if the DHOST theories involves a non-trivial $A_3$ (in which case the stealth SdS black holes also exist). Since the background geometry is exactly SdS, the scalar perturbation does not couple with the odd-parity metric perturbation. We also find that the even-parity perturbations is different from that in GR in general. The metric perturbation could be source the perturbation of external matter field in a different way due to the presence of the scalar field. More concerning,  we find that the scalar fluctuation sees a singular effective metric and hence suffers from a strong coupling problem. The black hole solution considered is therefore beyond the regime of validity of the DHOST effective field theory and cannot be trusted.\\

Finally we point out that the issue of strong coupling derived in this manuscript is very generic to a large class of  DHOST and Horndeski solutions. In particular those issues apply to other rotating black hole solutions with scalar hair found in the literature and in some sub-classes of Horndeski theories. We show that under a set of conditions DHOST and Horndeski solutions suffer the same scalar strong coupling issue
irrespectively of the specific manifold and symmetry of the system.

\acknowledgments

We would like to thank Christos Charmousis, Hayato Motohashi and Huan Yang for helpful discussion and suggestions.
CdR would like thank the Perimeter Institute for Theoretical Physics for its hospitality during part of this work and for support from the Simons Emmy Noether program. The work of CdR is supported by an STFC grant ST/P000762/1. CdR thanks the Royal Society for support at ICL through a Wolfson Research Merit Award. CdR and JZ are supported by the European Union's Horizon 2020 Research Council grant 724659 MassiveCosmo ERC-2016-COG. CdR is also supported by a Simons Foundation award ID 555326 under the Simons Foundation's Origins of the Universe initiative, `\textit{Cosmology Beyond Einstein's Theory}'.

\appendix

\section{Even-Parity Coordinate Transformations}\label{app:diff}

In this appendix we shall see the effect if an even-parity coordinate transformation so as to motivate our gauge chose in the study of the even sector of section \ref{sec:evenpert}.
Recalling that the even parity metric perturbations can be written as
\begin{equation}
h_{ab}^{\text{even}} = \begin{pmatrix}
A H_0 & \hspace{0.2cm}& H_1  & \hspace{0.2cm}& \mathcal{H}_0 \partial_\theta & \hspace{0.2cm}&  \mathcal{H}_0 \partial_\phi \\
H_1 & \hspace{0.2cm}& H_2/A  & \hspace{0.2cm}& \mathcal{H}_1 \partial_\theta & \hspace{0.2cm}& \mathcal{H}_1 \partial_\phi \\
\mathcal{H}_0 \partial_\theta & \hspace{0.2cm}& \mathcal{H}_1 \partial_\theta & \hspace{0.2cm}& \mathcal{K} + \G \nabla_\theta \nabla_\theta & \hspace{0.2cm}& \G \nabla_\theta \nabla_\phi\\
\mathcal{H}_0 \partial_\phi & \hspace{0.2cm}& \mathcal{H}_1 \partial_\phi & \hspace{0.2cm}& \G \nabla_\phi \nabla_\theta & \hspace{0.2cm}& \sin^2\theta \mathcal{K}+ \G \nabla_\phi \nabla_\phi
\end{pmatrix} Y_{\ell m} \, , \label{eq:appevenhab}
\end{equation}
and the scalar perturbation as
\ba
\delta \varphi = \Phi\left(t,r\right) Y_{\ell m},
\ea
we now consider an infinitesimal coordinate transformation $x^a \to \tilde x^a=x^a + \xi^a$ with
\ba\label{even diff}
	\xi^a = \left(\tT(t,r),\, \tR(t,r),\, \Theta (t,r) \partial_\theta,\, \frac{\Theta (t,r) \partial_\phi}{\sin^2 \theta} \right) Y_{\ell m} (\theta, \phi)\,.
\ea
Then the metric perturbations in Eq.~\eqref{eq:appevenhab} transform as follows:
\ba
\label{tildeH0}
H_0& \to & \tilde H_0  =  H_0 +  2 \dot \tT + \tfrac{A_{,r}}{A} \tR \\
H_1& \to & \tilde H_1  =  H_1 + A \tT_{,r}  - \dot \tR /A\\
H_2 & \to & \tilde H_2  =  H_2 + \tfrac{A_{,r}}{A} \tR - 2 \tR_{,r} \\
{\mathcal{H}}_0 & \to & \tilde{\mathcal{H}}_0  =  \mathcal{H}_0 + A \tT -  r^2 \dot \Theta \\
{\mathcal{H}}_1 & \to & \tilde{\mathcal{H}}_1  =  \mathcal{H}_1 - \tR/A - r^2 \Theta_{,r} \\
\G & \to & \tilde \G  =  \G - 2 \Theta\\
\mathcal{K} & \to & \tilde{\mathcal{K}}  =  \mathcal{K} -  \tfrac{2}{r} \tR\,.
\ea

For multipoles $\ell \ge 2$, one can set the gauge $\tilde \G=\tilde \K= \tilde{\mathcal{H}}_0  =0$ by an appropriate choice of the respective functions $\Theta, \tR$ and $\tT$ and can omit the tildes from now on.

For monopole, Eq.~\eqref{eq:appevenhab} becomes
\begin{equation}
\left.h_{ab}^{\text{even}}\right|_{\ell=0} = \frac{1}{2\sqrt{\pi}} \begin{pmatrix}
A H_0 & \hspace{0.2cm}& H_1  & \hspace{0.2cm}& 0 & \hspace{0.2cm}&  0 \\
H_1 & \hspace{0.2cm}& H_2/A  & \hspace{0.2cm}& 0 & \hspace{0.2cm}&  0 \\
0 & \hspace{0.2cm}& 0 & \hspace{0.2cm}& \mathcal{K}  & \hspace{0.2cm}& 0\\
0 & \hspace{0.2cm}& 0 & \hspace{0.2cm}& 0 & \hspace{0.2cm}& \sin^2\theta \mathcal{K}
\end{pmatrix}  \, ,
\end{equation}
while the gauge transformation \eqref{even diff} involves two free functions $T$ and $\tR$, which can be chosen appropriately so as to fix $\K = H_1 =0$.

Finally, for dipole, Eq.~\eqref{eq:appevenhab} becomes
\begin{equation}
\left.h_{ab}^{\text{even}}\right|_{\ell=1}  = \begin{pmatrix}
A H_0 & \hspace{0.2cm}& H_1  & \hspace{0.2cm}& \mathcal{H}_0 \partial_\theta & \hspace{0.2cm}&  \mathcal{H}_0 \partial_\phi \\
H_1 & \hspace{0.2cm}& H_2/A  & \hspace{0.2cm}& \mathcal{H}_1 \partial_\theta & \hspace{0.2cm}& \mathcal{H}_1 \partial_\phi \\
\mathcal{H}_0 \partial_\theta & \hspace{0.2cm}& \mathcal{H}_1 \partial_\theta & \hspace{0.2cm}& \mathcal{K} - \G  & \hspace{0.2cm}& 0\\
\mathcal{H}_0 \partial_\phi & \hspace{0.2cm}& \mathcal{H}_1 \partial_\phi & \hspace{0.2cm}& 0 & \hspace{0.2cm}& \sin^2\theta \left(\mathcal{K} -\G\right)
\end{pmatrix} Y_{1 m} \, , \label{eq:appevenhab}
\end{equation}
which depends on $\K$ and $\G$ only through $\K-\G$. The gauge transformation \eqref{even diff} still involves three free function $\Theta$, $T$ and $\tR$, which can be chosen so as to set $\K-\G = \scH_0=H_2=0$.

\section{Expressions of the Coefficients}\label{app}

The coefficients in Eq.~\eqref{eq:chiL} are defined as follow.
\ba
&&c_1 = \frac{(4+J-6A-6\Lambda r^2)A r}{1+J-3A-3\Lambda r^2} \\
&& c_2 = -\frac{\left[3A^2-6\Lambda A r^2 + (1+J-3\Lambda r^2)(3\Lambda r^2-1)\right]A^2 r}{1+J-3A-3\Lambda r^2} \\
&& c_3 = -\frac{\left[J^2+3A^2-2A(1+J+3\Lambda r^2)-(1-3\Lambda r^2)^2\right]A}{2r(1+J-3A-3\Lambda r^2)}\\
&&c_4 = J A^2 \\
&& c_5= \frac{JA(J-2A+rA_{,r})}{2r}
\ea
\ba
d_1= \frac12 A r^2, \quad d_2 = \frac12 A^3r^2, \quad d_3=\frac12 A^2, \quad d_4= - A r^2
\ea
\ba
&&f_1 = \frac{A\left[3A^2 + A(J-2-6\Lambda r^2) - (J+1-3\Lambda r^2)(1-3\Lambda r^2)\right]}{r(J+1-3A-3\Lambda r^2)} \\
&& f_2 = \frac{(J-2)J A}{r^2(J+1-3A-3\Lambda r^2)}
\ea
where $J = \ell (\ell+1)$.

\section{Horndeski Black Hole Solutions}
\label{app:HorndeskiBH}

Perturbations about static Black Hole solutions in Horndeski \eqref{eq:Horndeski} were explored in \cite{Kobayashi:2014wsa}. While the analysis performed in the manuscript applied to DHOST theories, one can show that they would be applicable to the special sub-class of solutions explored in \cite{Kobayashi:2014wsa} when $q=0$ and when taking $G_{3,X}(X_0)=K_{,X}(X_0)=0$, while keeping $G_4$ to be constant, $G_4(X)=\bar G_{4}=$const  and imposing $G_5$ to vanish identically.

Upon these restrictions, one can confirm that the variable $\Sigma$ defined in eq.~(36) of \cite{Kobayashi:2014wsa} vanishes  and the variable $\mathcal{P}_1$ defined in eq.~(34) is given by $\mathcal{P}_1=\bar G_{4}=\frac 12 \mathcal{F}$, hence implying that the dynamical metric $\mathcal{K}$ is always singular for that sub-class of solutions ${\rm det}(\mathcal{K})=0$ as can be seen from eq.~(38) of \cite{Kobayashi:2014wsa}, in agreement with the results presented here.

\bibliographystyle{JHEP}
\bibliography{master}
\end{document}